\documentclass{rmf-d}
\usepackage{nopageno,rmfbib,multicol,times,epsf,amsmath,amssymb,cite}
\usepackage[latin1]{inputenc}
\usepackage[]{caption2}
\usepackage{graphicx}
\def\rmfcornisa{Research OR Education of Physics \hfill\rmf\ {\bf ??} (*?*) ???--???
\hfill MES? A\~NO?}

\def\rmfcintilla{{\it Rev.\ Mex.\ Fis.\/} {\bf ??} (*?*) (????) ???--???}
\clearpage \rmfcaptionstyle 
\begin{document}
\title{   Hadron structure at small-$x$ via unintegrated gluon densities
\vspace{-6pt}}
\author{Andr\`ee Dafne Bolognino} 
\address{Dipartimento di Fisica, Universit\`a della Calabria \\ and Istituto Nazionale di Fisica Nucleare, Gruppo collegato di Cosenza, \\ I-87036 Arcavacata di Rende, Cosenza, Italy\\
ad.bolognino@unical.it}
\author{Francesco Giovanni Celiberto}
\address{European Centre for Theoretical Studies in Nuclear Physics and Related Areas (ECT*), I-38123
Villazzano, Trento, Italy \\
Fondazione Bruno Kessler (FBK), I-38123 Povo, Trento, Italy\\
INFN-TIFPA Trento Institute of Fundamental Physics and Applications, I-38123 Povo, Trento, Italy\\
fceliberto@ectstar.eu}
\author{Michael Fucilla}
\address{Dipartimento di Fisica, Universit\`a della Calabria \\ and Istituto Nazionale di Fisica Nucleare, Gruppo collegato di Cosenza, \\ I-87036 Arcavacata di Rende, Cosenza, Italy\\
michael.fucilla@unical.it}
\author{Dimtry Yu. Ivanov}
\address{Sobolev Institute of Mathematics, 630090 Novosibirsk, Russia \\
d-ivanov@math.nsc.ru}
\author{Alessandro Papa}
\address{Dipartimento di Fisica, Universit\`a della Calabria \\ and Istituto Nazionale di Fisica Nucleare, Gruppo collegato di Cosenza, \\ I-87036 Arcavacata di Rende, Cosenza, Italy\\
alessandro.papa@fis.unical.it}
\author{Wolfgang Sch\"{a}fer}
\address{Institute of Nuclear Physics Polish Academy of Sciences
ul. Radzikowskiego 152, PL-32-342, Krak\'ow, Poland \\
wolfgang.schafer@ifj.edu.pl}
\author{Antoni Szczurek}
\address{College of Natural Sciences, Institute of Physics, University of Rzesz\'ow \\
ul. Pigonia 1, PL-35-310 Rzesz\'ow, Poland \\
Institute of Nuclear Physics Polish Academy of Sciences
ul. Radzikowskiego 152, PL-32-342, Krak\'ow, Poland \\
antoni.szczurek@ifj.edu.pl}
\maketitle
\recibido{day month year}{day month year
\vspace{-12pt}}
\begin{abstract}
\vspace{1em} Inclusive as well as exclusive emissions in forward and central directions of rapidity are widely recognized as excellent channels to access the proton structure at small $x$. In this regime, to describe nucleons structure, it is necessary to use $k_T$-unintegrated distributions. In particular, at large transverse momenta, the $x$-evolution of the so-called unintegrated gluon distribution is driven by the Balitsky-Fadin-Kuraev-Lipatov equation, within the framework of the high-energy factorization (HEF). \\
Recent analyses on the diffractive electroproduction of $\rho$ mesons have corroborated the underlying assumption that the small-size dipole scattering mechanism is at work, thus validating the use of the HEF formalism. Nonetheless, a significant sensitivity of polarized cross sections to intermediate values of the meson transverse momenta, where, in the case of inclusive emissions, a description at the hand of the transverse momentum dependent (TMD) factorization starts to be the most appropriate framework, has been observed. \\
In this work, we will review the formal description of the UGD within the BFKL approach and present some UGD models that have been proposed, then we will describe the state of the art of some recent phenomenological analyses.  \vspace{1em}
\end{abstract}
\keys{  Hadron structure, Quantum Chromodynamics at small $x$   \vspace{-4pt}}
\pacs{\bf{\textit{12.38-t}}, \; \bf{\textit{12.38Cy}}    \vspace{-4pt}}

\begin{multicols}{2}

\section{Introduction}

After almost fifty years since Quantum Chromodynamics (QCD) was proposed, and subsequently recognized as the theory of the interaction between quarks and gluons, a correct and complete description of the proton structure, in terms of its constituents, remains a challenge for the scientific community. Examples, which allows to understand the complexity of such a description, are the mass-gap and the proton spin puzzle problems. Apart from the purely theoretical interest in the problem, the description of the hadronic structure is essential to understand many of modern problems of the Standard Model (SM). As we know, in fact, many of modern colliders, such as the Large Hadron Collider (LHC) and the Electron-Ion Collider (EIC), use proton or nuclei (as projectile and/or target); therefore an accurate description of the proton structure is essential 
to unveil signals of physics beyond the Standard Model. \\
In understanding the behavior of a hadron in high-energy collisions, we are always faced with the need to describe both perturbative and non-perturbative aspects of QCD. Thanks to the factorization theorems, we are often able to separate the two dynamics, in such a way as to be able to apply the computation techniques of perturbative quantum field theory to the so-called hard parts of processes and to reabsorb the part concerning the non-perturbative dynamics in some parton densities. Among these densities, the most general one is the Wigner distribution, which is a 5-dimensional object depending on longitudinal fraction of momenta ($x$), transverse momentum ($\vec{k}_T$) and transverse position ($\vec{b}_T$) of the parton inside the hadron. Integrating over some of these variables we obtain: 1) Generalized Parton Distributions (GPDs), if one integrates over transverse momenta, 2) Transverse Momentum Distributions (TMDs), if one integrates over transverse position, 3) Parton Distribution Functions (PDFs), if one integrates over both transverse momentum and position. Distributions have a fixed value (at fixed parameters), only at tree level; while, when considering quantum corrections, they become running quantities. Their evolution with respect to a scale can be described perturbatively, but their value at a given energy scale must be extracted through a fit on data. In the case of PDFs, universality properties guarantee that we can use PDFs extracted using a particular process, to predict cross sections of other processes. For more complicated objects, like TMDs, the situation is not so simple and we often talk about modified universality. It is also important to mention that each of these distributions satisfies different evolution equations.\\
A very important line of research concerns the behavior of parton densities (and especially of the unpolarized gluon one) in the so-called semi-hard regime of QCD, in which $s \gg \{Q^2\} \gg \Lambda_{QCD}$, where $s$ is the squared center-of-mass energy, $\{ Q^2 \}$ a set of typical hard scales and $\Lambda_{QCD}$ the QCD mass scale. In this regime, terms containing logarithms (of the form $\ln (Q^2/s)= \ln(1/x)$) become even more important than the collinear ones resumed to all orders through the Dokshitzer-Gribov-Lipatov-Altarelli-Parisi (DGLAP) evolution equations\cite{Dokshitzer:1977sg,Altarelli:1977zs}. 
To realistically describe the structure of the proton, we must introduce a $\vec{k}_T$ unintegrated gluon density (UGD), whose evolution at small-$x$ is governed by the Balitsky-Fadin-Kuraev-Lipatov (BFKL) equation\cite{Fadin:1975cb,Kuraev:1976ge,Kuraev:1977fs,Balitsky:1978ic}. BFKL became famous owing to the prediction of the rapid growth of the $\gamma^{*}$-$p$ cross section at increasing energy,
subsequently discovered experimentally. Therefore, the BFKL equation is usually associated with the evolution of the unintegrated gluon distribution, even if its applicability area is much wider. \\ 
Differently from collinear PDFs, the UGD
is not well known and several types of models for it do exist, which lead to very different shapes in the $(x, \vec{k}_T)$-plane. In the following, we use the $\rho$-meson leptoproduction to discriminate among different models of UGD.

\section{Models of unintegrated gluon density}

We start by giving the original definition of UGD in terms of the BFKL Green function. Let's consider the total $\gamma^{*}$-$p$ cross section in $k_T$-factorization:
\begin{equation}
    \begin{split}
    \sigma_{\lambda} (x, Q^2) = & \frac{\mathcal{G}}{(2 \pi)^4} \int \frac{d^2 \vec{k}_1}{\vec{k}_1^{2}} \\ & \times \int \frac{d^2 \vec{k}_2}{\vec{k}_2^{2}} \Phi_{\lambda} (\vec{k}_1) \Phi_{p}  (\vec{k}_2) F(x ,\vec{k}_1, \vec{k}_2) \; ,
    \end{split}
\end{equation}
where $\lambda$ is the virtual photon polarization, $\mathcal{G}$ incorporates color constants, $\Phi_{\lambda}$ and $\Phi_{p}$ are respectively the photon and proton impact factor and
\begin{equation}
\begin{split}
     F(x ,\vec{k}_1, \vec{k}_2) = \sum_{n=0}^{\infty} \int d \nu \left( \frac{\vec{k}_1^{2}}{\vec{k}_2^{2}} \right)^{i \nu} \frac{e^{i n (\theta_1-\theta_2)} e^{\bar{\alpha}_s \chi (n, \nu) \ln \left( \frac{1}{x} \right)}}{2 \pi^2 |\vec{k}_1||\vec{k}_2|}
\end{split}     
\end{equation}
is the Green function. The $x$-dependence of our final result comes from this object:
\begin{equation}
    F \sim \frac{x^{-\omega_0}}{\sqrt{\ln \left( \frac{1}{x} \right)}} \; , \hspace{0.5 cm} \omega_0 = 4 \alpha_s \ln 2 \; .
\end{equation}
We define the \textbf{unintegrated gluon density}, $\mathcal{F}(x, \vec{k})$, as
\begin{equation}
    \mathcal{F}(x, \vec{k}) \equiv \frac{1}{(2 \pi)^3} \int \frac{d^2 \vec{k}'}{\vec{k}'^2} \Phi_{p} (\vec{k}') \; \vec{k}^2 \; F(x,\vec{k}, \vec{k}') \; .
\end{equation}
Then, it is clear that
\begin{equation}
    \sigma_{\lambda} (x, Q^2) = \frac{\mathcal{G}}{(2 \pi)^4} \int \frac{d^2 \vec{k}_1}{(\vec{k}_1^2)^2} \Phi_{\lambda} (\vec{k}_1)  \mathcal{F} (x ,\vec{k}_1) \; .
    \label{BFKLCrossSection}
\end{equation}
$\mathcal{F}(x, \vec{k})$ is not a fully perturbative object, in fact it is expressed as the convolution of the BFKL Green function (known perturbatively) and a proton impact factor (non-perturbative object) that has to be modeled. Some modelizations, as we shall see, follow this scheme, while others parametrize directly $\mathcal{F}(x, \vec{k})$. \\
Models that will be considered in this work are\footnote{For reasons of space, we limit ourselves to mentioning the models used in this work. For a more complete review see~\cite{Bolognino:2021niq}}:
\begin{itemize}
    \item Gluon momentum derivative (toy model)
    \begin{equation}
        \mathcal{F}(x,k^2) = \frac{d (x g(x,k^2))}{d \ln k^2} \; .
    \end{equation}
    \item The \textbf{ABIPSW} model\cite{Anikin:2011sa}
    \begin{equation}
        \mathcal{F}(x,k^2) = \frac{A}{(2 \pi)^2 M^2} \frac{k^2}{k^2+M^2} \; .
    \end{equation}
    This is an $x$-independent model.
    \item The \textbf{IN} model\cite{Ivanov:2000cm} 
    \begin{equation}
        \mathcal{F}(x,k^2) = \mathcal{F}_{{\rm{soft}}}(x,k^2)+\mathcal{F}_{{\rm{hard}}}(x,k^2) \; .
    \end{equation}
    This is a soft-hard model developed with the purpose of probing different regions of transverse momentum.
    \item The \textbf{HSS} model \cite{Hentschinski:2012kr}
    \begin{equation}
        \mathcal{F}(x,k^2) = F(x, \vec{k}, \vec{k}') \otimes \Phi_p (\vec{k}') \; ,
    \end{equation}
    $\otimes$ means convolution in the transverse momenta $\vec{k}'$. This model is exactly based on the costruction of the UGD shown above.
    \item The \textbf{GBW} model \cite{Golec-Biernat:1998zce} \\
    This UGD parametrization derives from the effective dipole cross section $ \sigma (x, r)$ for the scattering of a $q \bar{q}$ pair off a nucleon.
    \item The \textbf{WMR} model \cite{Watt:2003mx} \\ 
    The unintegrated parton distributions are determined by imposing angular-ordering constraints on gluon emission. This UGD model satisfy the famous Catani-Ciafaloni-Fiorani-Marchesini (CCFM) evolution equation\cite{Ciafaloni:1987ur,Catani:1989sg,Marchesini:1994wr}.
    \item The \textbf{BCRT} model \cite{Bacchetta:2021oht} \\
    It is a small-$x$ improved model for the unpolarized gluon TMD based on the quark spectator model idea.
\end{itemize}
\section{$\rho$-meson leptoproduction: Theoretical set-up}
The second ingredient needed to build our prediction is the impact factor for the transition $\gamma_{\lambda} \rightarrow \rho_{\lambda'}$. We want to resolve this process in helicity, hence we will need more then one impact factor. Dominant helicity amplitudes are those corresponding to $\gamma_L \rightarrow \rho_L$ and $\gamma_T \rightarrow \rho_T$ transitions. Labelling by $T_{\lambda_{\gamma} \lambda'_{\rho}}$ the helicity amplitudes one has the following hierarchy:
\begin{equation}
    T_{00} \gg T_{11} \gg T_{10} \gg T_{01} \gg T_{-11} \; .
\end{equation}
In general, the impact factor for the photoproduction of a $\rho$-meson is a convolution between an hard coefficient function and a distribution amplitude (DA). In \cite{Anikin:2009bf} this object is expressed through a twist-expansion, achived by Taylor expanding the hard part. At leading twist we have two parton correlators connecting the hard and the soft part, at the next-to-leading twist we have an additional gluon, and so on. \\
\begin{itemize}
    \item $\gamma_{L} \rightarrow \rho_{L}$ impact factor \vspace{0.1 cm} \\
    This impact factor starts at the leading twist (twist two) and it is known up to next-to-leading order. The LO expression is \cite{Bolognino:2021niq}
    \begin{equation}
    \begin{split}
        \Phi_{\gamma_{L} \rightarrow \rho_{L}} & (k, Q; \mu^2) = 2 B \frac{\sqrt{N_c^2-1}}{Q N_c} \\ & \times \int_0^1 dy \varphi_1 (y; \mu^2) \left( \frac{\alpha}{\alpha + y \bar{y}} \right) \; ,
    \end{split}
    \end{equation}
    where $\alpha = k^2 / Q^2$, $B= 2 \pi \alpha_s \frac{e}{\sqrt{2}} f_{\rho}$ and 
    \begin{equation}
        \varphi_1 (y; \mu^2) = 6 y \bar{y} \left( 1 + a_2 (\mu^2) \frac{3}{2} (5(y-\bar{y})^2-1) \right)
    \end{equation}
     is the twist-2 DA.
    \item $\gamma_{T} \rightarrow \rho_{T}$ impact factor \vspace{0.1 cm} \\
    This impact factor starts at the next-to-leading twist (twist three) and it is known up to leading order. Its expression can be found in \cite{Bolognino:2018rhb}.
\end{itemize}
\section{$\rho$-meson leptoproduction: Phenomenology}
\begin{figure*}
\centering
\includegraphics[width=0.485\textwidth]{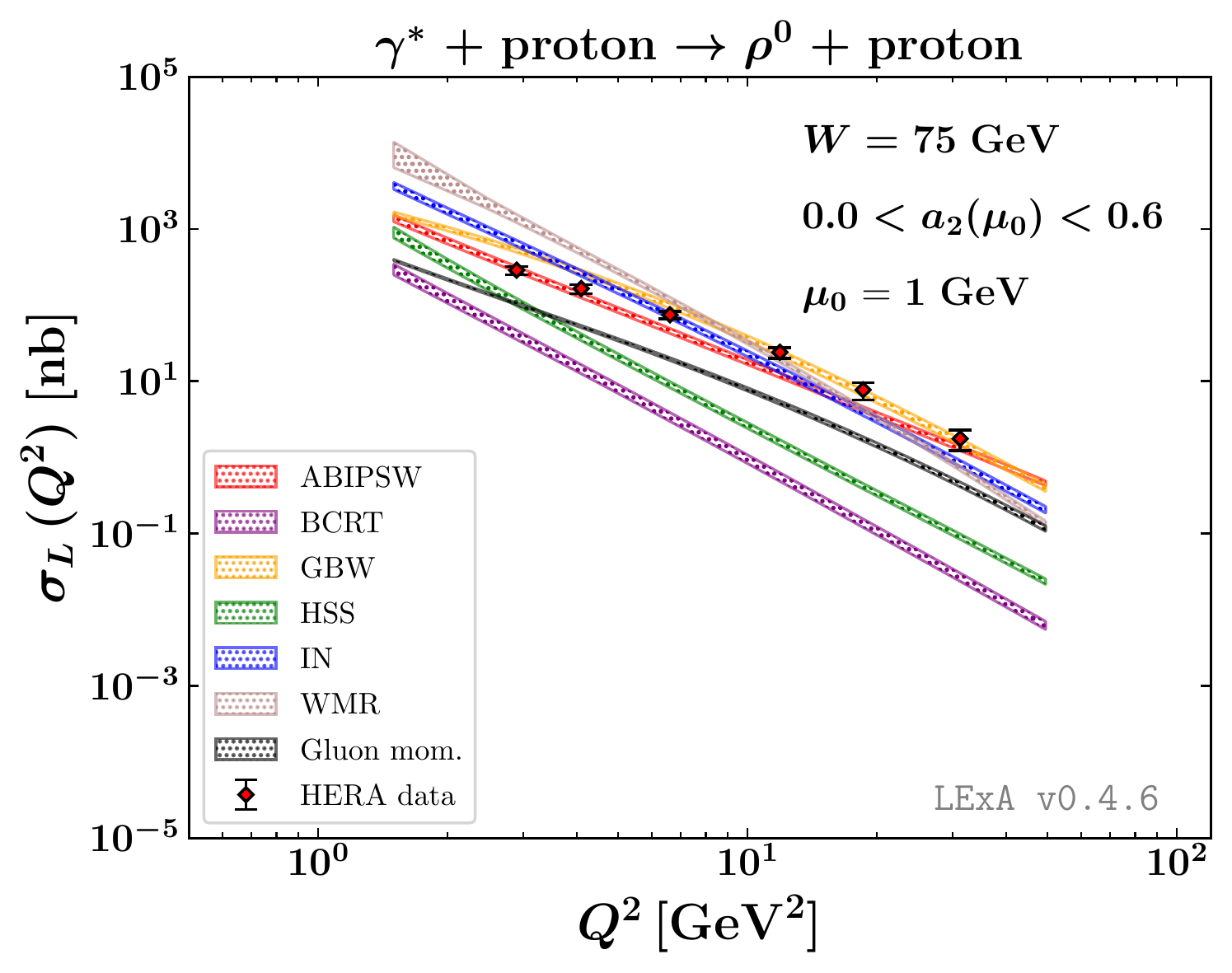}
\includegraphics[width=0.485\textwidth]{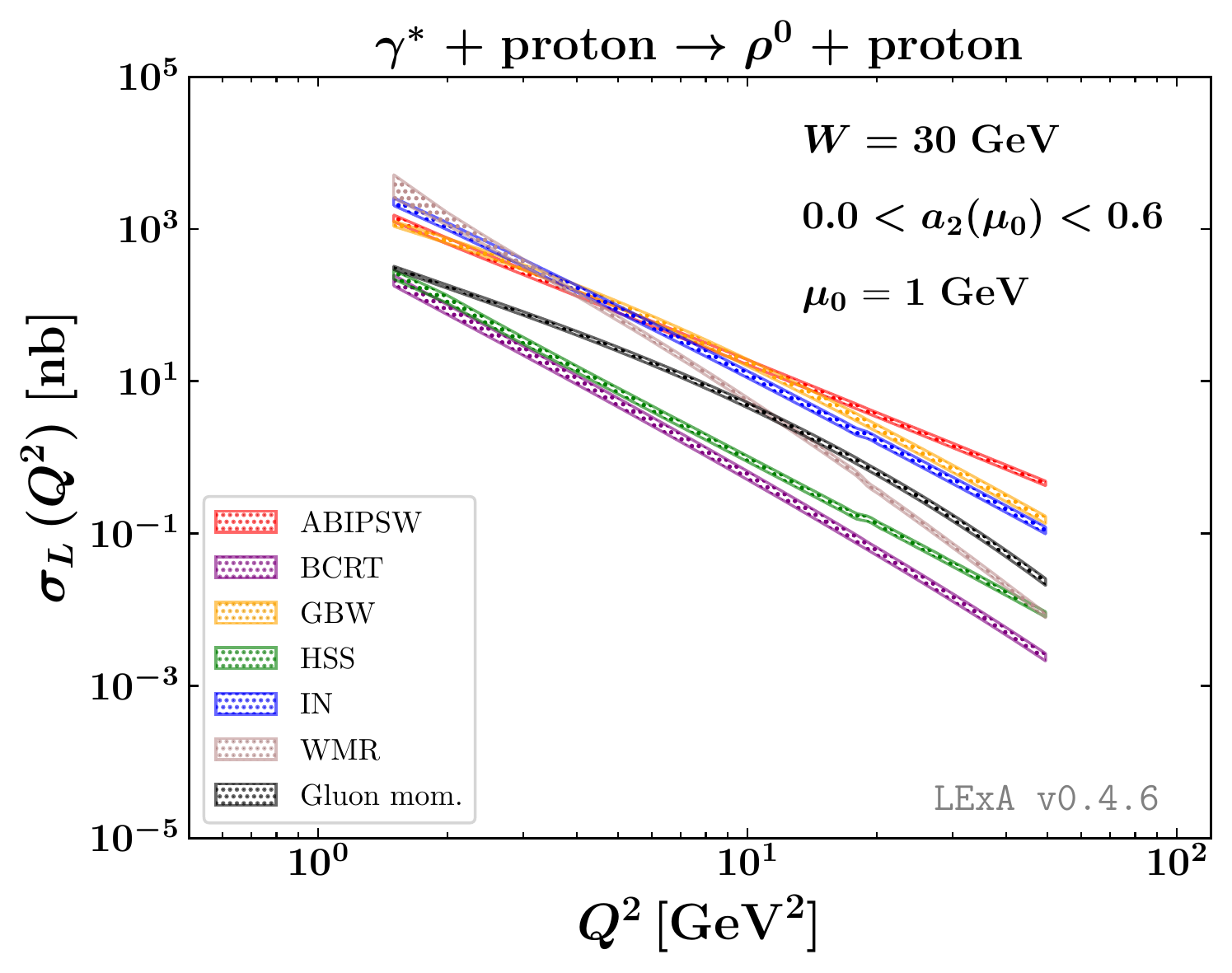}
\caption{Longitudinal cross-section as a function of the squared photon virtuality, $Q^2$, at $W = 75$ GeV$^2$ (left) and at $W = 30$ GeV$^2$ (right).}
  \label{sigmaL}
\end{figure*}

\begin{figure*}
\centering
\includegraphics[width=0.485\textwidth]{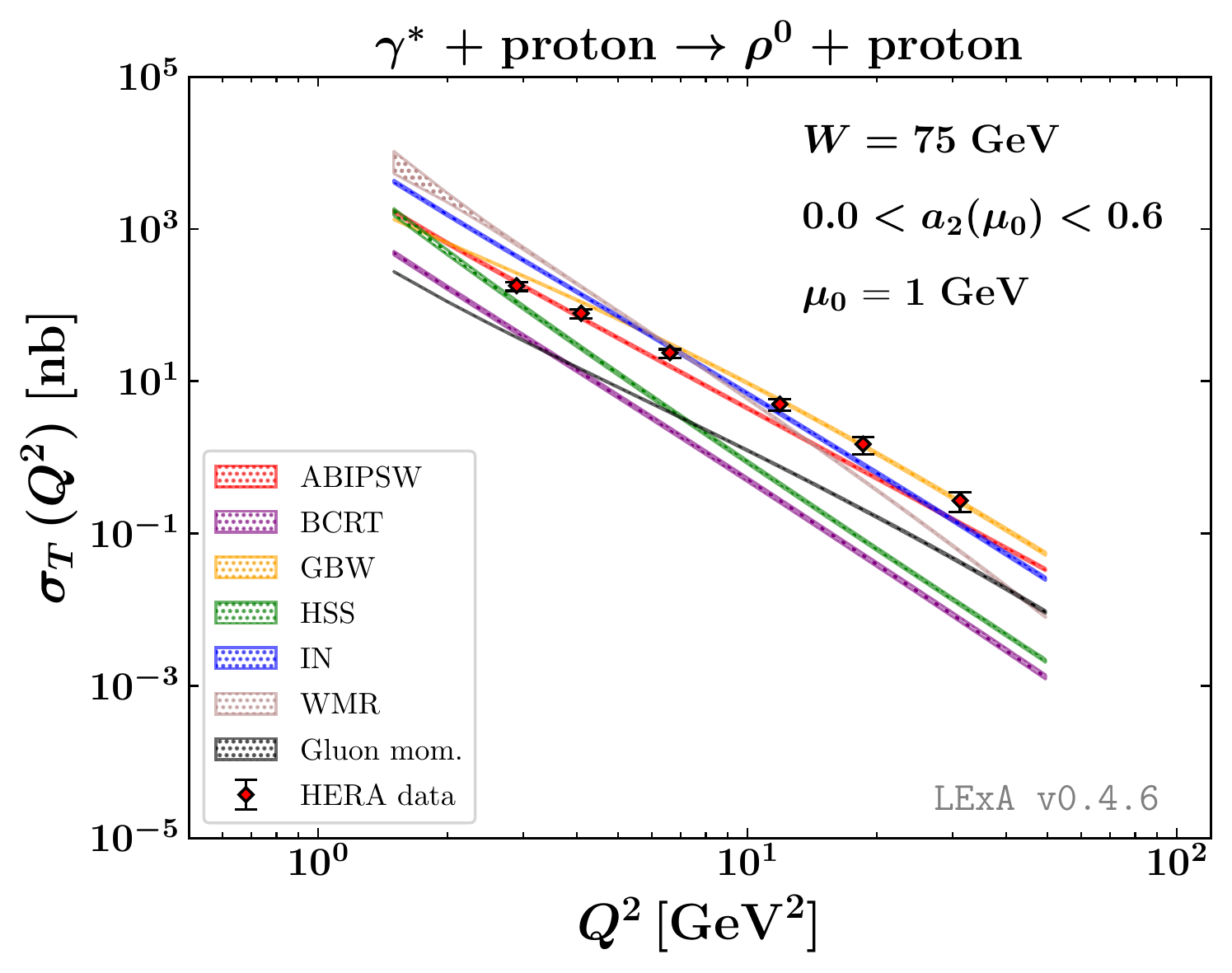}
\includegraphics[width=0.485\textwidth]{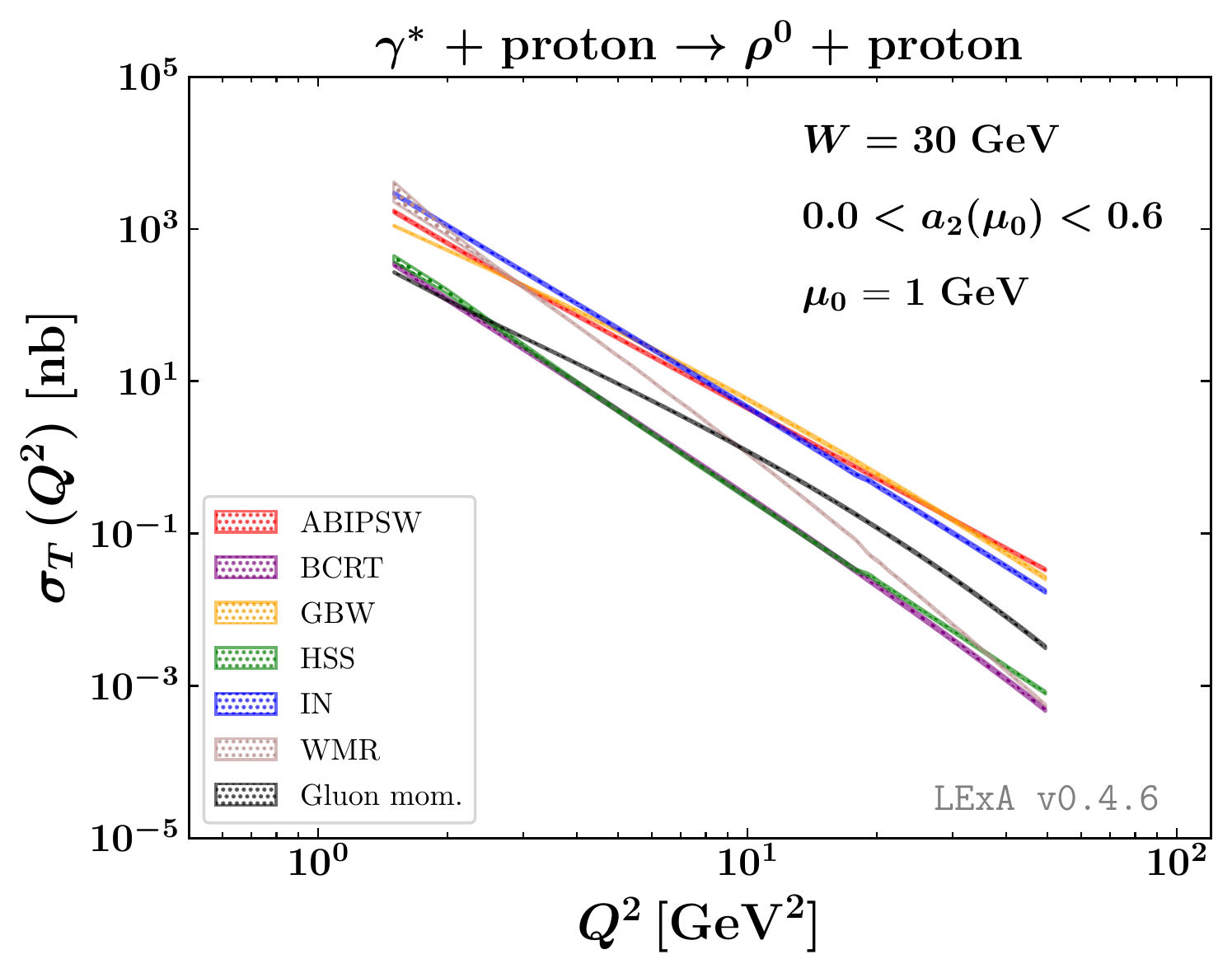}
\caption{Transverse cross-section as a function of the squared photon virtuality, $Q^2$, at $W = 75$ GeV$^2$ (left) and at $W = 30$ GeV$^2$ (right).}
  \label{sigmaT}
\end{figure*}

In this section we present some predictions for the $\rho$-meson leptoproduction in kinematical conditions typical of HERA collider and of the future EIC collider. In the case of HERA we show and discuss comparison with data. \\
The $T_{\lambda_{\gamma} \lambda'_{\rho}}$ can be expressed as
\begin{equation}
    T_{\lambda_{\gamma} \lambda'_{\rho}} = is \int \frac{d^2 k}{(k^2)^2} \Phi^{\gamma^{*}(\lambda_{\gamma}) \rightarrow \rho (\lambda'_{\rho})} \mathcal{F} (x, k^2) \; ,
\end{equation}
where $x=Q^2/s$. We stress that the amplitude in the semi-hard regime is dominated by its imaginary part. In the so-called Wandzura-Wilczek (WW) approximation, in which genuine terms are neglected, we have (for the two leading amplitudes):
\begin{equation}
\begin{split}
    T_{11} = is\frac{2BC}{Q^2} & \int \frac{d^2k}{(k^2)^2} \mathcal{F} (x, \vec{k}^2) \\ & \times \int  \frac{dy}{y \bar{y}} \frac{\alpha (\alpha+2 y \bar{y})}{(\alpha+ y \bar{y})^2} \varphi_+^{\rm WW} (y, \mu^2)\;,
    \end{split}
    \label{Amplitude1}
\end{equation}
\begin{equation}
\begin{split}
    T_{00} = is\frac{4BC}{Q^2} & \int \frac{d^2k}{(k^2)^2} \mathcal{F} (x, \vec{k}^2) \\ & \times \int dy \frac{\alpha}{(\alpha+y \bar{y})} \varphi_{+}^{\rm as} (y, \mu^2)\;,
    \end{split}
    \label{Amplitude2}
\end{equation}
where $C = \sqrt{4 \pi \alpha_{\rm em}}$. The expression of DAs $\varphi_+^{\rm WW} (y, \mu^2)$, $\varphi_{+}^{\rm as} (y, \mu^2)$ can be found in \cite{Bolognino:2019pba}. In the same work also the generalization of formulas (\ref{Amplitude1}) and (\ref{Amplitude2}) to massive quark case is given.
We will investigate the longitudinal and the transverse cross section,
\begin{equation}
\begin{split}
    & \sigma_{L} (\gamma^* p) = \frac{1}{16 \pi b(Q^2)} \frac{|T_{00}(s,t=0)|^2}{W^2} \;,\\ & \sigma_{T} (\gamma^* p) = \frac{1}{16 \pi b(Q^2)} \frac{|T_{11}(s,t=0)|^2}{W^2} \;.
    \hspace{0.5 cm}
    \end{split}
\end{equation}
Here $W^2 \equiv s$ and 
\begin{equation}
    b(Q^2) = \beta_0 - \beta_1 \ln \left[ \frac{Q^2 + m_V^2}{ m_{J/\Psi}} \right] + \frac{\beta_2}{Q^2 + m_V^2} 
\end{equation}
is the slope function for light vector mesons. For $\rho$-meson we have 
\begin{equation}
  \beta_0 = 6.5 \; {\rm{GeV}}^2, \; \beta_1 = 1.2 \; {\rm{GeV}}^2, \; \beta_2 = 1.1 \; {\rm{GeV^2}} \; .
\end{equation}
In Figure \ref{sigmaL} (\ref{sigmaT}) we show the longitudinal (transverse) cross section for the $\sigma_L$ ($\sigma_T$) as a function of the squared photon virtuality, $Q^2$; a more comprehensive phenomenological study can be found in~\cite{Bolognino:2021niq}. All cross sections fall at increasing $Q^2$ as it is expected from the high-energy analysis. Left panels refer to a typical HERA kinematics ($W=75$ GeV$^2$), while in the right panels there are predictions for measurements that can be made at the EIC. From comparison with data we can understand that none of the models is capable of describing the entire $Q^2$-spectrum of HERA. In particular, at small value of $Q^2$ (between 2 and 3 GeV$^2$) only the ABIPSW model fits data, while above 6 GeV$^2$ only the GBW model can describe data. As can be seen, these consideration remain true for both transverse and longitudinal cross sections. An intriguing possibility for future developments, motivated by the analysis presented above, is to consider a new UGD model, which contains a low-$k_T$ TMD input and encodes the small-$x$ evolution, to try to describe the entire HERA $Q^2$-spectrum.

\end{multicols}

\medline
\begin{multicols}{2}

\end{multicols}
\end{document}